\documentclass{emulateapj}

\shorttitle{Cosmic-ray Acceleration at the Shock front in IC1262}

\shortauthors{Hudson, and Henriksen}

\begin{document}

\title{Diffuse Non-thermal X-ray Emission: Evidence for Cosmic-ray Acceleration at the Shock front in IC1262}

\author{Daniel S. Hudson}
\affil{Physics Department, University of Maryland,
Baltimore County, 	Baltimore, MD 21250}

\author{Mark J. Henriksen\altaffilmark{1}}
\affil{Joint Center for Astrophysics, Physics Department, University of Maryland,
Baltimore County, 	Baltimore, MD 21250}

\altaffiltext{1}{Laboratory for High energy Astrophysics, NASA/GSFC}

\begin{abstract}

We report the first localization of diffuse, non-thermal, X-ray emission from a nearby galaxy cluster.  Using {\it Chandra} data, we have isolated a diffuse non-thermal X-ray component with a photon index, $\Gamma_{X}$ = 2.21$^{+0.14}_{-0.15}$ and a flux of 9.5$^{+1.1}_{-2.5} \times 10^{-5}$ photons cm$^{-2}$ s$^{-1}$ keV$^{-1}$ at 1 keV, that extends from $\sim$1$\arcmin$.5 to $\sim$2$\arcmin$.5 to the south of the X-ray flux peak.  Comparison to simulations implies that the diffuse non-thermal emission is produced by primary electrons, accelerated at shocks to relativistic velocities.  Using these results and the flux and hardness maps produced with data from the {\it Chandra} Advanced CCD Imaging Spectrometer, we conclude that a smaller subclump coming from the north merged with IC1262.  The offset of the cD galaxy from the X-ray peak and large peculiar velocity indicate that the subclump's impact parameter was to the west and on the near side of IC1262.

\end{abstract}

\keywords{galaxies: clusters: individual (IC1262) - galaxies:  intergalactic medium - galaxies: X-rays - galaxies: non-thermal emission}

\section{Introduction}

Recent {\it Chandra} observations indicate that relativistic plasma injected into the intracluster medium (ICM) from radio sources eventually detaches from the radio source, forming bubbles of radio plasma in the ICM \citep{mcnamara,heinz,fabian}.  In the model proposed by \citet{ensslin}, these radio ghosts survive in the ICM, and provide a seed population of Cosmic Rays(CRs).  Diffuse non-thermal (NT) emission is produced when merger induced shocks re-accelerate, via the first order Fermi process, this seed CR population.  Current evidence suggests that these radio ghosts contain a significant population of protons as well as electrons \citep{ensslin}.  Since \citet{Blasi} demonstrated that diffuse NT X-ray emission could be produced by either primary electrons directly accelerated at shock fronts, or secondary electrons produced during proton-proton collisions, there are two possible sources for the observed diffuse NT emission.  To determine the mechanism that produces diffuse NT emission requires accurate measurement of the spectrum and location of the NT emission.

Simulations by \citet{miniati} show that diffuse NT emission occurs in a broad mass range of clusters with a luminosity proportional to the X-ray temperature, making cool clusters and groups an important diagnostic for understanding which population of electrons produces diffuse NT emission.  They find that spectral index of the diffuse NT emission is dependent on the electron population producing the emission, such that the spectral index of diffuse NT emission produced from primary electrons has a steep spectral index ($\alpha_{ic}>$1.1), while for secondary it is systematically flatter ($\alpha_{ic}<$1.1) . 

\citet{hudson} reported detection of diffuse NT X-ray and radio emission from IC1262, a poor cluster of galaxies.  The X-ray detection was made using the {\it BeppoSAX} Medium Energy Concentrator Spectrometer (MECS) detector, and the radio using the NRAO VLA Sky Survey (NVSS) and Westerbork Northern Sky Survey (WENSS).  Although the MECS was able to constrain the region of the NT emission, it does not have the spatial resolution to identify the source of the NT emission.  In this paper, we show that the {\it Chandra} Advanced CCD Imaging Spectrometer (ACIS)  has the energy range and spatial capability to detect, localize, and characterize the spectrum of NT X-ray emission from low temperature clusters. These results are compared to simulations of cosmic-ray acceleration at shock fronts.

Throughout this letter, we assume a Hubble Constant of H$_{0}$ = 65 {\it h$_{65}$} km s$^{-1}$ Mpc$^{-1}$ and q$_{0}$ = $\case{1}{2}$.  Quoted confidence intervals are at a 90\% level, unless otherwise specified.

\section{Observations and Methods}
\begin{figure}[t]
\includegraphics[scale=.4,angle=0]{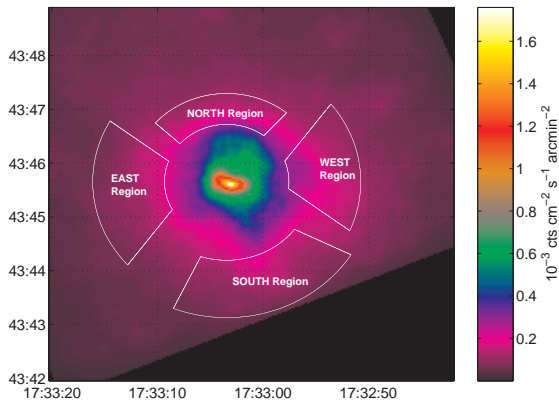}
\figcaption[f1.eps]
{Chandra {\it ACIS} Flux Map of IC1262.  The white figures are regions extracted to search for possible non-thermal emission. \label{fig1}}
\end{figure}

The IC1262 galaxy cluster is a poor cluster of galaxies located at (J2000) 17$^{h}$ 33$^{m}$ 01.0$^{s}$, +43$^{\circ}$ 45$\arcmin$ 28$\arcsec$ \citep{ebeling} with  a redshift of 0.0343 \citep{colles}, so that 1$\arcmin$ = 46 h$_{65}^{-1}$ kpc.  It was observed by the {\it Chandra} ACIS S3 chip on 23 August 2001 for $\sim$ 31 ksec.  The total count rate of the uncleaned data is 10.2 $\pm$ 0.02 cts s$^{-1}$ with a peak of 37 $\pm$ 3 cts s$^{-1}$, which is well below the saturation limit of the ACIS chips operating in Faint Mode (170 cts s$^{-1}$).  Figure-\ref{fig1} is a color coded intensity map that shows the full 8$\arcmin.5 \times$ 8$\arcmin$.5 image in the 0.3-8.0 keV band.  The image was obtained using processing routines outlined in the CIAO 2.3 Science Threads\footnote{{\tt http://cxc.harvard.edu/ciao/threads/index.html}}.  The data was CTI corrected and cleaned for flares, point sources, and anomalous high background. Exposure maps were created for 17 different energy bands to ensure proper exposure correction.  Background was taken from the CALDB 2.21 blank-sky datasets.

\begin{figure}[t]
\includegraphics[scale=.4,angle=0]{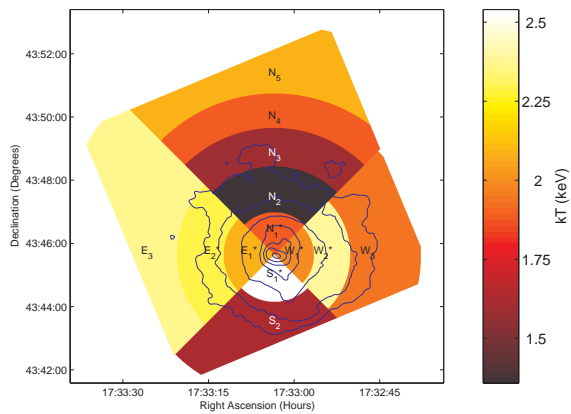}
\figcaption[f3.eps]
{Temperature Map of IC1262.  Regions with a (*) are have Temperatures given for the double thermal model.\label{fig3}}
\end{figure}

To construct the temperature map (see Figure-\ref{fig3}) we divided the chip into quadrants, north, south, east, and west.  Spectra were extracted from pie shaped regions with radii differences such that there were at least 20000 Counts in each region.  In order to ensure proper background subtraction, we eliminated periods of high background, following the {\it Filtering Lightcurves} thread\footnote{{\tt http://cxc.harvard.edu/ciao/threads/filter\_ltcrv/}}.  In order to account for background variability, the background was normalized to the source in the 10-12 keV range \citep{markevitch}.  The data were grouped so that there were at least 30 counts per channel.  All models included either an {\it acisabs} component or had an {\it acisabs} corrected ancillary response file in order to account for the time dependent absorption (at low energies) of the ACIS window.  Results are given in Table-\ref{tbl-1}.

\begin{deluxetable*}{lcccccc}
\tabletypesize{\scriptsize}
\tablewidth{0pt}
\tablecolumns{6}
\tablecaption{Spectral Fits of Regions. \label{tbl-1}}
\tablehead{\colhead{Region} & \colhead{Model\tablenotemark{\dag}} & \colhead{kT$_{1}$ (keV)} & \colhead{Abundance} & \colhead{kT$_{2}$ (keV)} & \colhead{$\chi^{2}$/dof}}
\startdata

N$_{1}$ & Apec		& 1.27$^{+0.03}_{-0.03}$ & 0.14$^{+0.03}_{-0.03}$ & 				& 173.0422/98 \\
N$_{1}$ & Apec+Apec 	& 1.88$^{+0.17}_{-0.15}$ & 0.44$^{+0.17}_{-0.14}$ & 0.83 $^{+0.16}_{-0.04}$ 	& 134.0729/96 \\
N$_{2}$ & Apec 		& 1.34$^{+0.12}_{-0.04}$ & 0.14$^{+0.05}_{-0.03}$ & 				& 119.7284/91 \\
N$_{3}$ & Apec 		& 1.57$^{+0.13}_{-0.16}$ & 0.12$^{+0.07}_{-0.05}$ & 				& 57.09452/76 \\
N$_{4}$ & Apec 		& 1.88$^{+0.43}_{-0.37}$ & 0.13$^{+0.14}_{-0.08}$ & 				& 65.19046/59 \\
N$_{5}$ & Apec 		& 2.09$^{+0.95}_{-0.58}$ & 0.12$^{+0.34}_{-0.12}$ & 				& 55.56707/48 \\
\\
W$_{1}$ & Apec 		& 1.80$^{+0.09}_{-0.12}$ & 0.25$^{+0.06}_{-0.05}$ & 				& 126.3223/108 \\
W$_{1}$ & Apec+Apec 	& 1.98$^{+0.12}_{-0.13}$ & 0.38$^{+0.11}_{-0.09}$ & 0.61$^{+0.20}_{-0.24}$ 	& 113.5917/106 \\
W$_{2}$ & Apec 		& 2.09$^{+0.18}_{-0.12}$ & 0.33$^{+0.11}_{-0.08}$ & 				& 136.9889/101 \\
W$_{2}$ & Apec+Apec 	& 2.38$^{+0.22}_{-0.21}$ & 0.57$^{+0.25}_{-0.16}$ & 0.28$^{+0.09}_{-0.08}$ 	& 123.7163/99 \\
W$_{3}$ & Apec 		& 1.93$^{+0.16}_{-0.16}$ & 0.18$^{+0.07}_{-0.06}$ & 				& 118.0369/107 \\
\\
S$_{1}$ & Apec 		& 1.62$^{+0.05}_{-0.05}$ & 0.27$^{+0.05}_{-0.04}$ & 				& 203.8204/103 \\
S$_{1}$ & Apec+Apec 	& 2.54$^{+0.26}_{-0.22}$ & 0.63$^{+0.13}_{-0.22}$ & 1.05$^{+0.21}_{-0.07}$ 	& 147.7749/101 \\
S$_{2}$ & Apec 		& 1.60$^{+0.08}_{-0.10}$ & 0.17$^{+0.06}_{-0.05}$ & 				& 134.1819/97 \\
\\
E$_{1}$ & Apec 		& 1.62$^{+0.05}_{-0.06}$ & 0.22$^{+0.05}_{-0.04}$ & 				& 114.6884/101 \\
E$_{1}$ & Apec+Apec 	& 2.08$^{+0.32}_{-0.16}$ & 0.49$^{+0.28}_{-0.13}$ & 0.82$^{+0.21}_{-0.08}$ 	& 87.91611/99 \\
E$_{2}$ & Apec		& 2.02$^{+0.21}_{-0.18}$ & 0.23$^{+0.11}_{-0.08}$ &				& 90.03336/89 \\
E$_{2}$ & Apec+Apec	& 2.30$^{+0.29}_{-0.36}$ & 0.40$^{+0.23}_{-0.20}$ & 0.23$^{+0.09 }_{0.21}$	& 83.43614/87 \\
E$_{3}$ & Apec		& 2.37$^{+0.37}_{-0.37}$ & 0.39$^{+0.25}_{-0.17}$ &				& 109.8093/99 \\

\enddata
\tablenotetext{\dag}{All models include photoelectric absorption using Wisconsin cross-sections frozen at galactic value of 2.46 $\times$ 10$^{20}$ cm$^{-2}$.}
\end{deluxetable*}

\section{Results and Analysis}

\begin{figure}[t]
\includegraphics[scale=.4,angle=0]{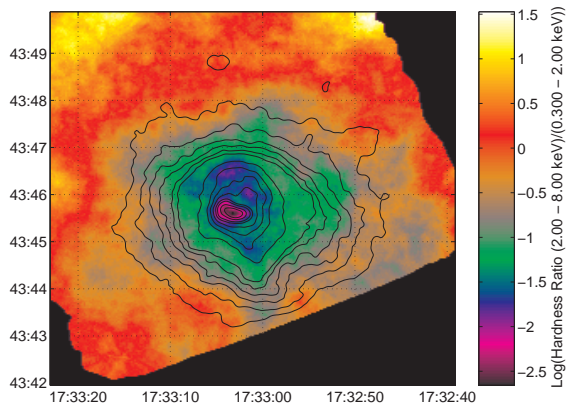}
\figcaption[f2.eps]
{Logscale Hardness Map of IC1262.  (2.0-8.0 keV)/(0.3-2.0keV) \label{fig2}}
\end{figure}

From the flux and hardness map, Figure-\ref{fig1} and Figure-\ref{fig2} respectively, we chose four regions that have a strong gradient in flux and hardness.  These regions are shown in Figure-\ref{fig1}.  We considered four models of background subtraction, to ensure any NT detection would not be the result of undersubtracted particle background.  These four background subtraction techniques were: (1) standard blank-sky fields extracted from the same region on the S3 Chip as the source and normalizing in the 10-12 keV range as described in \citet{markevitch}, (2) extracting a region far to the north of the peak flux, modeling it and then using the model times a free constant in our source + background spectrum modeling, (3) freezing the background model described in (2) with a constant proportional to the difference in size of the source and background region, and (4) subtracting the size normalized background spectrum from the source spectrum.  In all cases the source model parameters are consistent within 90\% confidence, indicating that all four methods produce the same results.  

\begin{figure}[t]
\includegraphics[scale=.17,angle=0]{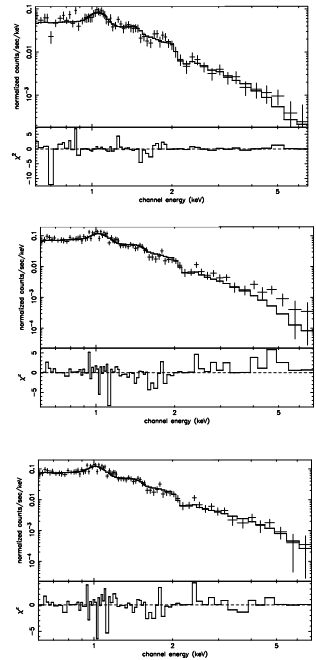}
\figcaption[f4.eps]
{Spectral fits to regions of compact isophotes:  top is single thermal fit to East Region, middle is single thermal fit to South Region, and bottom is thermal + powerlaw fit to the South Region. \label{fig4}}
\end{figure}

We modeled our four "possible shock" regions with a single thermal component model, a double component thermal model, and a thermal model with powerlaw.  The addition of the powerlaw component does not significantly improve the fit in the East and West Regions.  In addition, Figure-\ref{fig4} shows that a thermal model fits the entire spectrum in the East Region.  In the case of the North Region, adding either a second thermal component, or a power-law component improves the fit.  The single thermal model fit to the South Region has clear high end residuals (see Figure-\ref{fig4}).  Although the addition of a second  second thermal component improves the fit ($\Delta \chi^{2} \sim$ 16 for a reduction of 2 degrees of freedom(DOF)), the addition of a powerlaw component provides the best fit ($\Delta \chi^{2} \sim$ 23 for a reduction of 2 DOF) and reduces the residuals (see Figure \ref{fig4}).  Although we cannot rule out a thermal interpretation of the second component, we argue a non-thermal interpretation is better statistically and physically.   The addition of a powerlaw (37\% probability of exceeding $\chi^{2}$ = 74 for 71 DOF) produces a better fit than adding a second thermal component (18\% probability of exceeding $\chi^{2} = 81.8$ for the 71 DOF).  The 90\% lower confidence of $\sim$7 keV implies a shock with a Mach number of $\sim$3, which is higher than observed in clusters \citep[and references therein]{gabici}.

In order to confirm that the nonthermal detection was not simply undersubtracted particle background, we repeated our four different techniques for handling the background that we had done for a single thermal model.  As in the case with the single thermal model, the results are consistent within 90\% for all the background techniques used (see Table-\ref{tbl-2}).  Also, we point out that the lack of a NT detection in the West and East Regions provides more evidence that the detection in the North and South Regions is not simply particle background, since that would be observed isotropically across the detector.

\begin{deluxetable*}{lcccccc}
\tabletypesize{\scriptsize}
\tablewidth{0pt}
\tablecolumns{7}
\tablecaption{Nonthermal detection in the North and South regions. \label{tbl-2}}
\tablehead{\colhead{Region} & \colhead{Background Model} & \colhead{kT$_{1}$ (keV)} & \colhead{Abundance} &  \colhead{$\Gamma_{X}$} & \colhead{Non-thermal} & \colhead{$\chi^{2}$/dof}\\ & & & & &  \colhead{Normalization($\times 10^{-4}$)} & }
\startdata
North \\
 & blank-sky       & 1.32$^{+0.15}_{-0.12}$ & $>1.00^{+>0.0}_{-0.79}$ & 2.36$^{+0.14}_{-0.12}$ & 0.79$^{+0.09}_{-0.34}$ & 52.93678/50 \\
 & model-bkg       & 1.29$^{+0.16}_{-0.13}$ & $>1.00^{+>0.0}_{-0.73}$ & 2.30$^{+0.19}_{-0.17}$ & 0.78$^{+0.11}_{-0.35}$ & 54.05200/49 \\
 & model-bkg(Frzn) & 1.31$^{+0.17}_{-0.14}$ & 0.39$^{+>0.61}_{-0.15}$ & 2.41$^{+0.24}_{-0.45}$ & 0.56$^{+0.22}_{-0.35}$ & 57.22508/50 \\
 & -local region   & 1.30$^{+0.17}_{-0.14}$ & 0.43$^{+>0.57}_{-0.29}$ & 2.40$^{+0.23}_{-0.42}$ & 0.58$^{+0.21}_{-0.36}$	& 54.61599/50 \\
\\
South \\
 & blank-sky	   & 1.44$^{+0.18}_{-0.13}$ & $>1.00^{+>0.0}_{-0.62}$ & 2.21$^{+0.14}_{-0.15}$ & 0.95$^{+0.11}_{-0.25}$	& 74.33978/71 \\
 & model-bkg	   & 1.40$^{+0.18}_{-0.14}$ & $>1.00^{+>0.0}_{-0.62}$ & 2.14$^{+0.16}_{-0.17}$ & 0.91$^{+0.13}_{-0.27}$	& 77.66279/70 \\
 & model-bkg(Frzn) & 1.44$^{+0.20}_{-0.15}$ & 0.49$^{+>0.51}_{-0.32}$ & 2.33$^{+0.33}_{-0.51}$ & 0.59$^{+0.26}_{-0.42}$ & 90.72980/71 \\
 & -local region   & 1.43$^{+0.19}_{-0.14}$ & $>1.00^{+>0.0}_{-0.76}$ & 2.27$^{+0.18}_{-0.38}$ & 0.78$^{+0.11}_{-0.46}$ & 76.28895/71 \\
\enddata
\end{deluxetable*}

\begin{figure}[t]
\epsscale{1}
\plotone{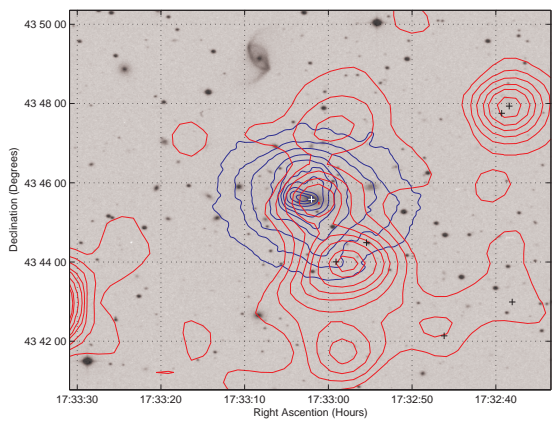}
\figcaption[f5.eps]
{Digital Sky Survey Image of IC1262 with {\it Chandra ACIS} X-ray flux contours (blue) and NVSS radio flux contours (red).  The X-ray contours are 0.1, 0.15, 0.2, 0.3, 0.4, 0.5, 0.6, 0.75, 1, and 1.3 $\times$ 10$^{-3}$ photons cm$^{-2}$ s$^{-1}$ arcmin$^{-2}$. Radio contours are 0.45 1.35, 2.70, 4.50, 6.30, 9.00, 13.50, and 15.75 mJy beam$^{-1}$. The black and white crosses are Faint Radio Sources at Twenty centimeters. \label{fig5}}

\end{figure}

The radio halo detected by \citet{hudson} appears to extend to the south along the elongated X-ray isophotes and through the South region (see Figure-\ref{fig5}).  However the $\sim$1$\arcmin$.0 resolution of the NVSS and WENSS precludes the removal of radio point sources.  We, therefore, consider two possibilities: (1) that the extended southern radio emission is dominated by radio point sources, and (2) that it is diffuse radio emission.  In the case that the radio emission is dominated by radio point sources, we argue that this would have no affect on our X-ray results.  Relativistic electrons diffusing from a point source at their Alfv\'{e}n speed, would only reach $\sim 5$ h$_{65}^{-1}$ kpc ($\sim  6 \arcsec$ ) (assuming a 1 $\mu$G magnetic field)  in their radiative lifetime ($\sim$10$^{8}$ years).  Therefore any radio point source with detectable X-ray emission would appear point-like and would have been removed from the {\it Chandra ACIS} data.  In the latter case, the diffuse radio emission is emitted by the same electrons producing the inverse compton emission (as in the inverse compton scattering-syntrotron model \citep{reph}).  We also note in passing that there are several identified radio point sources (including the cD galaxy) in IC1262, which could provide the seed population of CRs necessary in the model proposed by \citet{ensslin}.

\section{Discussion}
A comparison of Figure-\ref{fig1} and Figure-\ref{fig2} to flux and temperature maps created from simulations of cluster mergers done by \citet{roe} and \citet{takizawa}, indicates a recent ($\sim$0.25 - 1 Gyr ago) north-south merger of a smaller subclump (mass ratio $\sim$1:2 - 1:4) with IC1262.  These simulations show the general characteristics of the temperature and morphological structure: elongation of the isophotes to the south, compression to the north, and the hottest gas south of the center.  Since cD galaxies in relaxed systems will reside at the bottom of the potential well (e.g. \citet{oegerle}), the offset of cD galaxy from the X-ray peak and its high peculiar velocity ($\Delta$v = 453 km s$^{-1}$) provide additional evidence of a merger.  The kick to initiate sloshing is parallel to the impact parameter \citep{tittley}.  As the simulations indicate a recent collision, the subclump must have passed to the west, pulling the cD galaxy to the west.

The elongation of the X-ray isophotes toward the south that overlap the diffuse radio emission open up the possibility that the elongation is due to non-thermal emission rather than a merger.  However, the evidence for a merger comes not only from the similarity of the southern X-ray isophotes to those seen in the simulations cited above, but also from the compressed northern X-ray isophotes and the very hot region that is south of center.

Comparing our results with the simulations done by \citet{miniati}, we argue that the localized diffuse NT emission is from a population of primary electrons.  \citet{miniati} find that primary electron reacceleration follows the shock region closely.  In the central region (r $<$ 0.3 Mpc), they find the greatest difference between the spectral index produced from primary electrons ($\alpha_{ic} \geq$1.1) and secondary electrons ($\alpha_{ic}<$1.1).  Our 90\% confidence photon index $\Gamma_{X}$ = 2.07 - 2.36 which corresponds to a spectral index $\alpha_{ic}$ = 1.07 - 1.36 is consistent with the spectral index that \citet{miniati} found for primary electrons.  Using the best fit F$_{ic}$-T$_{x}$ of \citet{miniati}, we find a flux of (10 - 48) $\times$ 10$^{-13}$ ergs cm$^{-2}$ s$^{-1}$ over 0.13 to 100 keV for primary electrons in a 1.94 keV cluster (IC1262's temperature as determined by \citet{hudson}).  The range fluxes depends on the R$_{e/p}$, the ratio of electrons to protons at relativistic energies.  Using the same energy range and cluster temperature we find that secondary electrons produce a flux of 1.9 $\times$ 10$^{-13}$ ergs cm$^{-2}$ s$^{-1}$.  For our best fit photon index $\Gamma_{X}$ = 2.21 and 90\% confidence normalization range of 7.0-11.1 $\times$ 10$^{-5}$ photons cm$^{-2}$ s$^{-1}$ keV$^{-1}$ at 1 keV, we determine a flux of (6 - 10) $\times$ 10$^{-13}$ ergs cm$^{-2}$ s$^{-1}$ for the 0.13 - 100 keV range.  This result has a lower limit 3$\times$ greater than the results predicted for flux from secondary electrons.  Although the 90\% upper confidence limit is barely consistent with the primary electron model (suggesting R$_{e/p} \sim$ 0.01), we point out that the results of \citet{miniati} is for shocks over the entire cluster, and our detection is only for a fraction of the cluster.  That is the total non-thermal emission from cluster is probably greater than our detection, further strengthening the argument that the flux is too high to come from secondary electrons.




\end{document}